\documentclass[aps, twocolumn, prx,superscriptaddress]{revtex4-1}
\usepackage{graphicx}
\usepackage{dcolumn}
\usepackage{natbib}
\usepackage[caption=false]{subfig}
\usepackage{color}
\usepackage{ORI_Group_style}

\newcommand{\Olab}{O\bold{e}_x\bold{e}_y\bold{e}_z}

\newcommand{\Obod}{O\nn_1\nn_2\nn_3}

\newcommand{\eula}{\alpha}
\newcommand{\eulb}{\beta}
\newcommand{\eulc}{\gamma}

\newcommand{\deula}{\Dot{\alpha}}
\newcommand{\deulb}{\Dot{\beta}}
\newcommand{\deulc}{\Dot{\gamma}}

\newcommand{\ww}{\boldsymbol{\w}}
\newcommand{\NN}{\bold{N}}

\begin{document}

\title{Spin Read-out of the Motion of Levitated Electrically Rotated Diamonds}

\author{Maxime Perdriat}
\affiliation{Laboratoire De Physique de l'\'Ecole Normale Sup\'erieure, ENS, PSL, CNRS,
Sorbonne Université, Université de Paris, 24 rue Lhomond, 75005 Paris, France.}

\author{Cosimo C. Rusconi}
\affiliation{Munich Center for Quantum Science and Technology,Schellingstrasse 4, D-80799 München, Germany.}
\affiliation{Max-Planck-Institut für Quantenoptik, Hans-Kopfermann-Strasse 1, 85748 Garching, Germany.}

\author{Tom Delord}
\affiliation{Department of Physics, City College of the City University of New York, New York, NY 10031 USA.}

\author{Paul Huillery}
\affiliation{Univ Rennes, INSA Rennes, CNRS, Institut FOTON - UMR 6082, F-35000 Rennes, France.}

\author{Clément Pellet-Mary}
\affiliation{Department of Physics, University of Basel, CH-4056 Basel, Switzerland.}

\author{Benjamin A. Stickler}
\affiliation{University of Duisburg-Essen, Faculty of Physics, Lotharstraße 1, 47048 Duisburg, Germany.
and Institute for Complex Quantum Systems, Ulm University - Albert-Einstein-Allee 11, D-89069 Ulm, Germany
}

\author{Gabriel H\'etet$^1$}

%
\begin{abstract}
Recent advancements with trapped nano- and micro-particles have enabled the exploration of motional states on unprecedented scales. Rotational degrees of freedom stand out due to their intrinsic non-linearity and their coupling with internal spin degrees of freedom, opening up possibilities for gyroscopy and magnetometry applications and the creation of macroscopic quantum superpositions. However, current techniques for fast and reliable rotation of particles with internal spins face challenges, such as optical absorption and heating issues.  Here, to address this gap, we demonstrate electrically driven rotation of micro-particles levitating in Paul traps. We show that micro-particles can be set to rotate stably at 150,000 rpm by operating in a hitherto unexplored parametrically driven regime using the particle electric quadrupolar moment. Moreover, the spin states of nitrogen-vacancy centers in diamonds undergoing full rotation were successfully controlled, allowing accurate angular trajectory reconstruction and demonstrating high rotational stability over extended periods. These achievements mark progress toward interfacing full rotation with internal magnetic degrees of freedom in micron-scale objects. In particular, it extends significantly the type of particles that can be rotated, such as ferromagnets, which offers direct implications for the study of large gyromagnetic effects at the micro-scale.

\end{abstract}

\maketitle

In the last decade, advances in the control and levitation of nano- and micro-particles have paved the way for studying mechanical mode dynamics on unprecedented scales \cite{GonzalezBallestero2021,Kuhn2017,Ahn2018,Reimann2018,Rashid2018,Delord2020}. Recent achievements in ground state cooling of center of mass modes have even allowed exploration of mechanical dynamics in a quantum regime~\cite{delic2020cooling,magrini2021real,tebbenjohanns2021quantum,piotrowski2023simultaneous}. 
In this context, rotational degrees of freedom possess two distinctive features that make them particularly attractive for future investigations~\cite{Stickler2021,GonzalezBallestero2021,Perdriat2021}. 
First, the rotational motion is intrinsically non-linear even in the absence of an external potential. This non-linearity represents a resource which can be controlled to produce unique dynamical effects both in the classical and quantum regime~\cite{Stickler2018,Ma2020}.
Second, internal spin degrees of freedom naturally couple to the mechanical rotation of the hosting particle as a consequence of the Einstein--de Haas and Barnett effects~\cite{Einstein1915,Barnett1915} or through the magnetic torque in the presence of a magnetic field ~\cite{Delord2017,Delord2020}.
These spin-mechanical couplings significantly affect the dynamics of micro- and nano-particles~\cite{Delord2017,Rusconi2016,Rusconi2017,Rusconi2017PRB,Kustura2022,Sato2022}. It unlocks the possibility to control the rotation via the internal spins~\cite{Delord2020,Perdriat2021,Ma2021,Rusconi2022}, as well as to use levitated particles for applications in gyroscopy~\cite{Wood2018,Wood2020} and magnetometry~\cite{Kimball2016}. In the single spin limit, spin-mechanical coupling may serve to generate macroscopic quantum superpositions of rotation~\cite{Rusconi2022}. Precise control of the rotational degree of freedom of particles with internal spin degrees of freedom, and notably the ability to fully rotate them, is therefore highly desirable. However at present, reliable and non-invasive techniques for fast rotation of such particles are lacking. 

Recent demonstrations of full rotation of levitated particles are based on optical rotation of silica nanoparticles~\cite{Arita2013,Kuhn2015,*Kuhn2017,*Kuhn2017b,Zielinska2023}, which have been rotated with frequencies as high as a few GHz~\cite{Hoang2016,*Reimann2018,*Jin2021,*Ju2023}.
However, extending these methods to particles with an internal magnetic structure such as diamond or ferromagnets is challenging. 
This is mostly due to the large optical absorption of such crystalline particles and the resulting heating from the intense laser beam \cite{Hoang2016}.

In this article we bridge this gap by demonstrating a novel mechanism for stable and non-invasive rotation of levitating particles, and apply this method to distinct particle species such as silica rods and diamonds. Entering the nominally unstable regime for the angular motion of particles in standard Paul traps and exploiting the non-linearity of the angular motion, we discover that elongated particles can be set in rotation in the kHz frequency range. The rotational frequency of the particle is locked to the trap frequency, and can thus be tuned by controlling the latter. Additionally, this new rotating technique is based on the electric quadrupole moment of the particle, unlike rotation {\it via} an  electric field which requires the particle to carry an electric dipole~\cite{Kane2010,Coppock2016,Nagornykh2017}. Since the electric dipole moment tends to vanish for highly charged particles~\cite{Martinetz2021} and no additional RF electrodes are required, this rotating technique is more versatile.
Furthermore, we successfully controlled the spin state of NV centers in diamonds undergoing full rotation. We could then employ NV magnetometry to accurately reconstruct its angular trajectory and show a remarkable stability of the rotational mechanism over several hours. 
These demonstrations represent very first steps towards interfacing full rotation with internal magnetic degrees of freedom with a micron-scale object.

The article is structured as follows. In \secref{sec:Locking}, we present the experimental set-up and explain the physical origin of the rotating regime of particles in Paul traps. Using this technique, we rotate silica microrods, obtain an experimental stability diagram for the angular dynamics and present a theoretical model that explains the underlying mechanisms. In \secref{sec:Rotate_NV}, we employ the spin-resonance of NV centers in electrically rotated micro-diamonds to reconstruct the diamond angular trajectory and demonstrate the single-axis character of the rotation mechanism. 

%
\begin{figure*}[t]
\includegraphics[width=2\columnwidth]{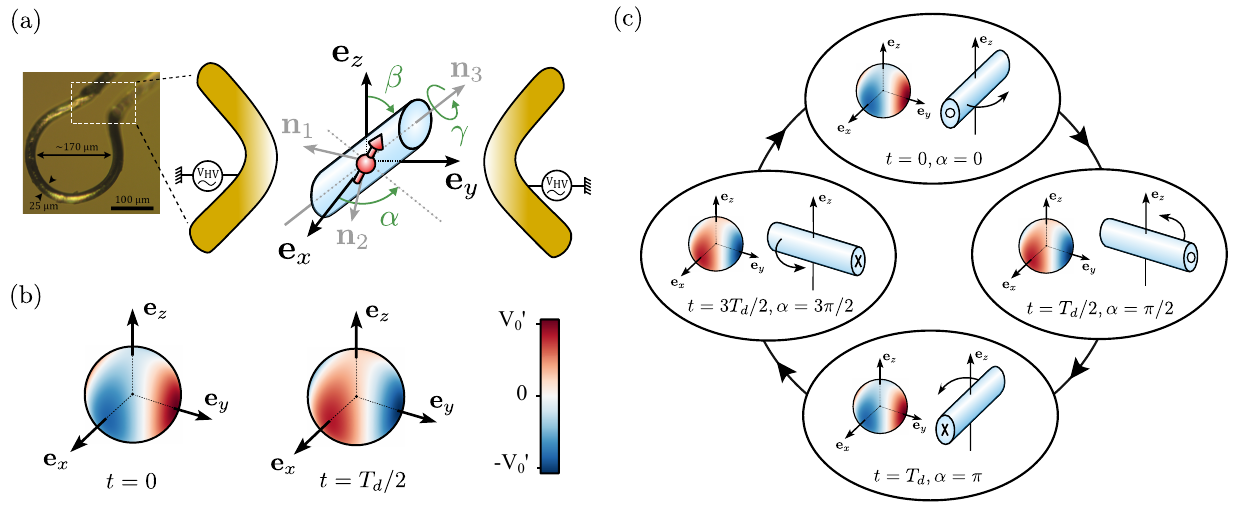}
\caption{(a) Left: photo of the Paul trap electrodes used in our experiments. The particle is trapped in the bottleneck region highlighted by the white box. Right: illustration of the system and definition of the orientation angle for a levitated asymmetrical cylinder. (b) Electric potential given in \eqnref{eq:V_p} created by the electrodes at two different times $t=0$ and $t=T_\text{Paul}$. $V_0'=V_0 (r_0/\ell_0)^2$. (c) Diagram explaining the rotational-locking effect.}\label{fig:Fig1}
\end{figure*}

\section{Electric rotational-Locking in a Paul Trap}\label{sec:Locking}

\subsection{Experimental Set-up} \label{sec:Experimental_Setup}

Our experimental platform consists in a bottleneck asymmetric Paul trap.
\figref{fig:Fig1}-(a) shows a picture as well as a sketch of the trap with an elongated particle represented by a deformed cylinder. We have also schematically included a spin inside the particle to represent the internal degree of freedom associated with the spins in diamond. The trap generates the time-dependent electric potential:
\be\label{eq:V_p}
    V(\rr,t) \equiv \frac{V(t)}{\ell_0^2}(a_x x^2+ a_y y^2 + a_z z^2).
\ee
Here, $V(t)=V_0\cos(\W_d t)$ is the electrode AC-voltage, $\ell_0$ the characteristic length scale of the trap, and $a_j$ the geometric parameters verifying $a_x+a_y+a_z=0$ and the  $a_x<a_z<0<a_y$. 
The trap operates at an electric potential $V_0 \approx 500$-$2000~\text{V}$ and a frequency $\W_d/2\pi \approx 1$-$10$~kHz. The length scale of the trap is typically $\ell_0 \approx 50~\mu{\rm m}$. It allows to levitate charged particles in the $10~\mu$m size range with thousands of charges on the surface \cite{Delord2016}.

Two different types of particles will be considered in this work: diamonds and silica micro-rods. 
Due to the quadrupolar form of the electrostatic potential in \eqnref{eq:V_p}, only the particle's monopole (total charge), dipole, and quadrupole moments in the particle electric multipolar expansion play a role in the dynamics. The monopole moment is responsible for a confining potential for the particle's center of mass. The dipole moment couples rotation and center of mass motion~\cite{Martinetz2021}. However, such coupling has never been observed in our experiments \cite{Delord2020}. Consequently, we neglect the contribution of the dipole moment for the remainder of the paper. Due to its quadrupole moment, the levitating particle is subjected to an oscillating electric torque. When this torque is averaged over one cycle of the Paul trap voltage, it can result in restoring torques for the three angular degree of freedom of the particle. As we will explore in detail in the following part, this oscillating torque can also lead to a more complex angular dynamics.

\subsection{The rotational-locking effect} \label{sec:Experimental_Setup}

Previous studies have already demonstrated the ability to completely confine the angular degrees of freedom of levitated particles using their quadrupole moment in a Paul trap~\cite{Delord2017,Delord2017APL}. We refer to this regime as the \emph{librating regime}. In this regime, the dynamic equations for the three Euler angles can be linearized around specific angular positions, resulting in three independent Mathieu equations (for more details, see Sec. \ref{app:Theory} in the appendix). The relevant parameter $q_u$ of the Mathieu equation can be computed based on the Paul trap and particle parameters. Stable librations are expected when $q_u \lesssim 0.9$ for all three angles, similar to the requirements for center of mass confinement in Paul traps. However, once $q_u \lesssim 0.9$ is no longer satisfied for at least one angle, that angle can become unstable and non-linearities in the angular electric potential can no longer be ignored. In such case, a different angular dynamics can emerge, the \emph{rotational-locking regime}. In this regime, the particle completes a full rotation about the $\mathbf{e}_z$ axis, with a rotation frequency half that of the AC voltage. The rotation period is simply twice the Paul trap period, given by $T_{\rm{rot}} = 2 / (\Omega_d/2\pi)$. 

We now derive the complete nonlinear dynamical equation for the angular dynamics. We neglect the center of mass to rotational coupling due to the small dipole moment (see above). Therefore, we can solely focus on the rotational dynamics of the system. The angular motion is ruled by the Euler equations,
\be\label{eq:Euler_EoM}
    \Dot{\LL} = \NN - \Gamma \LL,
\ee
where $\LL \equiv {\rm I} \ww$ is the angular momentum of the particle, with inertia tensor ${\rm I}$ and angular frequency vector $\ww$, $\NN$ is the electric torque exerted by the trapping potential, and $\Gamma$ the tensor representing the gas-induced damping \cite{Martinetz2018}.
The electric torque reads~\cite{Martinetz2021}
\be\label{eq:Torque}
    \NN = \frac{2 V(t)}{3\ell_0^2}\sum_{j=x,y,z} a_j \bold{e}_j \times Q(\W) \bold{e}_j,
\ee
where $Q(\W) \equiv R(\W) Q_0 R^T(\W)$ is the electric quadrupole moment in the laboratory frame. Its expression in the body-fixed frame reads $Q_0 \equiv \int \text{d}\rr\, \rho_0(\rr) \pare{3\rr\otimes\rr - r^2\id}$, where $\rho_0(\rr)$ denotes the charge distribution on the nanoparticle surface in its reference orientation \cite{Martinetz2021}.

Assuming that the particle long axis is kept fixed in the $xy$-plane, as will be discussed later, the reduced equation of the particle angular dynamics can be obtained. 
It then reads
\be\label{eq:Parmeteric_Pendulum}
	\Ddot{\eula} + \gamma_0 \Dot{\eula} + \w_0^2 \cos(\W_d t)\sin(2\eula)  =0.
\ee
Here $\eula$ describes the orientation of the particle's longest axis in the $xy$-plane [see \figref{fig:Fig1}-(a)], $\gamma_0/2\pi$ is the damping rate, and $\w_0^2 \equiv V_0(a_x-a_y)(Q_2-Q_3) /3\ell_0^2 I_1$, where $I_1$ is the moment of inertia for a rotation about a direction perpendicular to the elongated axis and $Q_i, i=\{1,2,3\}$ are the eigenvalues of the quadrupolar tensor.
\eqnref{eq:Parmeteric_Pendulum} describes a parametrically excited pendulum for which gravitational acceleration is neglected. The parametric pendulum is known to have two different types of stable solutions depending on the values of the frequency and amplitude of the parametric drive~\cite{Kapitza1951,VanDerWeele2001}. These solutions corresponds to (i) oscillatory motion about a fixed direction and (ii) rotational motion around its pivot~\cite{Xu2005,Litak2008}.
We expect \eqnref{eq:Parmeteric_Pendulum} to also exhibit these two kinds of solutions corresponding respectively to the librating and rotational-locking regimes of the particle dynamics.
Note that this analysis is valid only if the particule symmetry axis remains in the $xy$-plane at any given time. Physically, this condition is reached once the particle rotates, thanks to the gyroscopic effect that provides additional angular confinement to the $\eulb$ and $\eulc$ oscillations as shown by Eqs.~(\ref{eq:W_beta}-\ref{eq:W_gamma})  in \appref{app:Theory}. We will also confirm that this single rotation axis regime is fulfilled experimentally. 

\begin{figure*}[t]
	\includegraphics[width=2\columnwidth]{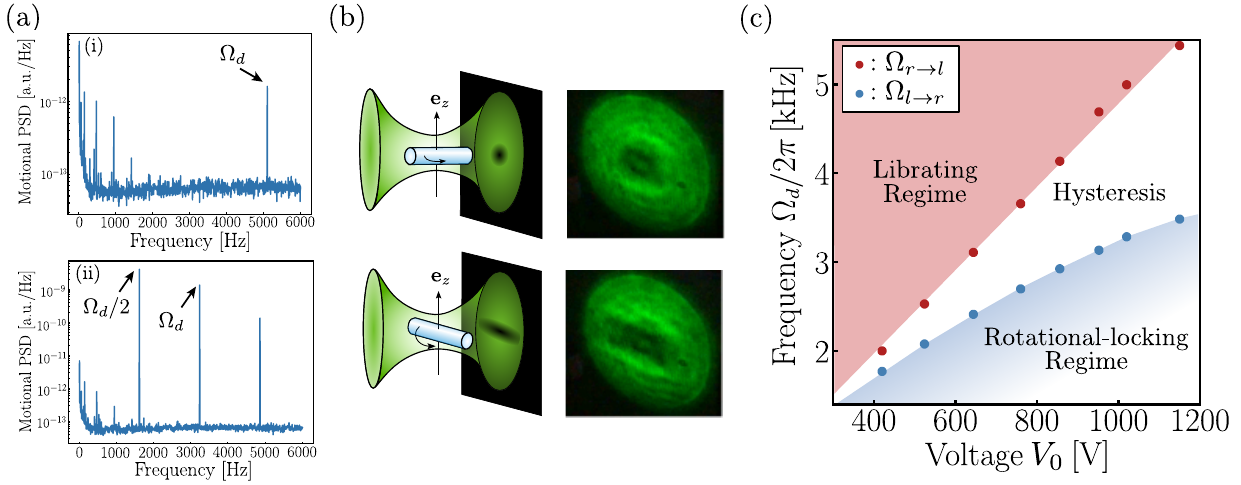}
	\caption{Experimental results. (a) (i) and (ii) : Power Spectral Densities of the angular motion of a silica micro-particle in the librating regime in (i) and in the rotational-locking regime in (ii). (b) Stroboscopic imaging of the rotation. Upper two panels show a schematic of a rotating rod at time $t=0$ (left panel) together with the corresponding shadow on the distant screen (right panel). The lower panels show the situation at $t=T_d/2$ where the rod is rotated by $\pi/2$. (c) Experimental reconstruction of the phase diagram showing the different dynamical regimes of the rod rotation. The red (blue) dots corresponds to the parameter for which the particle switches from the rotating (librating) regime to the librating (rotating) regime when the frequency is increased (decreased).}\label{fig:Fig2}
\end{figure*}

In order to provide a simple physical picture of the rotational-locking effect, we introduce the potential $V(\rr,t)$ generated by the trap on a spherical surface at a distance $r_0$ from the trap center. We consider a levitated particle that is positively and homogeneously charged. The potential is plotted in \figref{fig:Fig1}(b) at two different times $t=0$ (left) and $t=T_d/2$ (right).  The potential is minimum in the $x$-axis at time $t=0$, while it is minimum in the $y$-axis at time $t=T_d/2$.
A schematic explanation of the rotational-locking mechanism is then provided in  \figref{fig:Fig1}.(c). To distinguish between the two ends of the particle, a circle and a cross are drawn on the two opposite faces. When $t=0$, the particle preferentially aligns its long axis along the $x$-axis, which corresponds to the angular position $\alpha=0$ that minimizes the quadrupolar electric energy. When $t=T_d/2$, the electric potential minima are rotated by an angle $\alpha=\pi/2$, causing the particle to align along the $y$-axis. When $t=T_d$, the electric potential is equal to its initial value ($t=0$). Because of the angular inertia of the particle, it is energetically favorable for the particle to continue on its rotational motion and to reach the position $\alpha=\pi$ instead of $\alpha=0$. The particle finally completes a full rotation after two periods of the Paul trap drive. The particle rotational motion at $\W_d/2$ is thus parametrically sustained by the electric potential oscillation, similar to a Kapitza pendulum. Note that the particle could also rotate in the opposite direction.

\subsection{Rotational-Locking with silica micro-rods}\label{sec:Experimental_Results}

In this section, we show rotational-locking of levitated silica micro-rods and identify the parameters which allow stable libration and stable rotational-locking. 

We work with well calibrated silica micro-rods (from Nippon Electric Glass Company) with a diameter of $4~\mu{\rm m}$ and a length of $15~\mu{\rm m}$. Such a large aspect ratio enables straightforward angular motion visualization.
The loading of the particles is done at ambient pressure, similarly to in \cite{Delord2016}.
The particles are then illuminated by a green laser and their motion is detected by collecting a portion of the light transmitted by the particle, which is then sent to a photodetector. This detection technique is sensitive both to the center of mass and the angular motion. 
%
The signal from the detector is then sent to a spectrum analyzer to measure the power spectral density (PSD) of the motion. \figref{fig:Fig2}(a) shows two different PSD signals, that have been obtained from two different silica micro-rods at atmospheric pressure. 

A first class of PSD is shown in \figref{fig:Fig2}(a)-(i). There, a typical sharp peak at the frequency $\W_d/2\pi$ is obtained, which indicates excess micro-motion of the particle center of mass and angular degrees of freedom, as a result of a displacement of the particle away from the Paul trap potential minimum. The other peaks are detection artifacts, such as electronic noise of the detector. The particle center of mass and angular degrees of freedom are overdamped by gaz collisions, so the Brownian motion of the particle can be seen as a broad noise in the low-frequency range (up to $\approx 500$Hz). Such a signal indicates stable center of mass and angular motion.

The second class of PSD signals is presented in \figref{fig:Fig2}(a)-(ii) for a different Paul trap drive.  In this case, a sharp peak at the frequency $(\W_d/2)/2\pi$ is also present. The presence of this peak indicates that the particle undergoes parametric mechanical motion at half the drive frequency of the Paul trap, corresponding to the expected rotational frequency in the rotational-locking regime. 
To confirm that this peak corresponds to full rotation, we employ stroboscopic measurements using an Acousto-Optic Modulator (AOM). The AOM generates short laser pulses at a slightly detuned frequency from $(\W_d/2)/2\pi$. This technique allows us to observe the particle in slow motion on a distant screen. \figref{fig:Fig2}(b) shows the shadow observed on a screen when illuminating a particle with a pulsed laser, with main particle axis pointing along the optical axis (left) or when it is rotated by $\pi/2$ (right).
The observation of the parametric motion at the frequency $(\W_d/2)/2\pi$ confirms the complete rotational motion of the particle around the $\mathbf{e}_z$ axis, as anticipated. 

The analysis of the PSD, along with the stroboscopic detection, provides us with a simple and robust experimental method to quickly determine the angular dynamics of the levitating particle.
We can thus proceed to explore the different angular regimes as a function of the drive frequency $\W_d/2\pi$ and amplitude $V_0$. Specifically, we start from a particle initially in the librating regime, and we decrease the drive frequency, while keeping the voltage $V_0$ fixed. The drive frequency is decreased until the only stable regime is the rotational-locking regime. We then increase the drive frequency to come back to the librating region. This protocol is performed for different values of the trap voltage $V_0$ with the same silica micro-rod. The results are presented in a dynamical phase diagram shown in \figref{fig:Fig2}.(c). We observe that for a given $V_0$ the value $\W_{l\rightarrow r}$ at which the particle switches from the librating to the rotational-locking regime is always lower than the value $\W_{r\rightarrow l}$ at which the particle returns to the librating regime from the rotational-locking regime. The two frequencies $\W_{l\rightarrow r}$ and $\W_{r\rightarrow l}$ thus define a \emph{hysteresis region} in the dynamical behaviour of the system when $V_0$ is varied.

\subsection{Theoretical analysis of the transition between the librating and rotational-locking regimes}\label{sec:Theory}

\begin{figure*}
	\includegraphics[width=2\columnwidth]{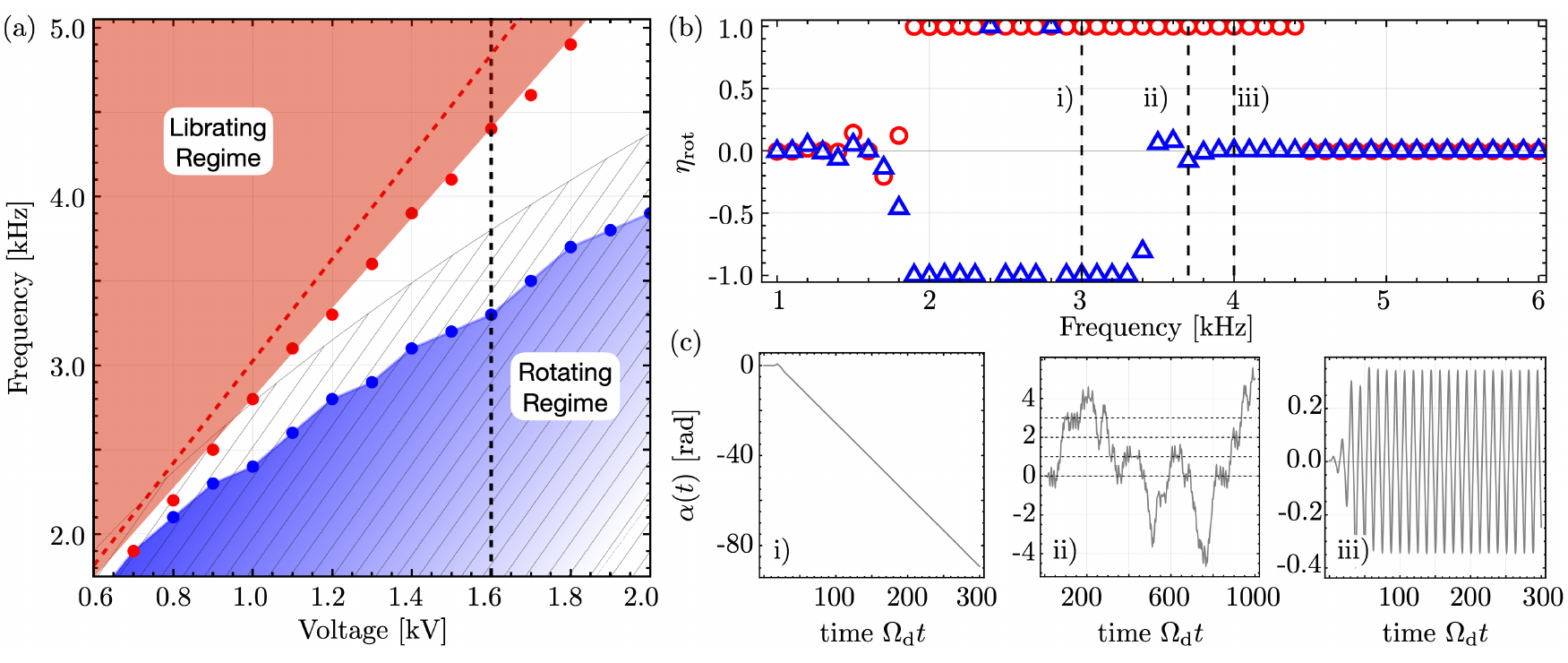}
	\caption{(a) Dynamical stability diagram predicted theoretically from \eqnref{eq:Parmeteric_Pendulum}. The red (blue) circles correspond to the transition frequency $\W_{r\rightarrow l}$ ($\W_{l\rightarrow r}$) from the rotating (librating) regime to the librating (rotating) regime. The red (blue) colored region denote the regime where only libration (locked rotation) is stable. The white region is the hysteresis region. The hatched area indicates the unstable region of \eqnref{eq:Parmeteric_Pendulum}. The black dotted vertical line corresponds to the value of $V_0$ for which we compute $\eta_\text{rot}$ as shown in panel (b). (b) Variation of $\eta_\text{rot}$ as a function of the driving frequency for a particle initially in the librating (red circles) and rotating (blue triangles) regimes. We assumed the initial conditions $\{\eula(0)=0.01, \Dot{\eula}(0)=0.01~\W_d/2\}$, and $\{\eula(0)=5\times10^{-3}, \Dot{\eula}(0)=1.01~\W_d/2\}$ for the librating and rotating regime respectively. (c) Three examples of trajectories for a particle initially in the librating regime for three different values of $\W_d$ as specified by the corresponding label i), ii), and iii) in panel (b). Specifically (from left to right) $\W_d/2\pi=3.0~\text{kHz},3.7~\text{kHz}$, and $4.0~\text{kHz}$. In this plot, we assumed $\gamma_0/2\pi = 1~\text{kHz}$.}\label{fig:Fig3}
\end{figure*}

The librating and rotational-locking regimes can be identified using the value of the following order parameter 

\be\label{eq:frot}
	\eta_\text{rot} \equiv \frac{2}{T\W_d}\int_{t_0}^{t_0+T}\!\!\! \text{d}\tau\, \Dot{\eula}(\tau).
\ee
It represents the averaged angular velocity over a time interval $T$, starting from a time $t_0$ which should be chosen such as to avoid any initial transient dynamics. In particular, $\eta_\text{rot} = 0$ corresponds to the oscillatory motion of the particle in the librating regime, while $\eta_\text{rot} = \pm 1$ describes locked clockwise and counter-clockwise rotation in the rotating regime.

In the large driving frequency limit, the numerical solutions of \eqnref{eq:Parmeteric_Pendulum} show small oscillations around the equilibrium orientations $\eula= k\pi/2$, $\Dot{\eula}=0~\text{rad.s}^{-1}$, where $ k \in \mathbb{Z}$, so that $\eta_\text{rot}=0$ in this regime. This is in agreement with what we observe experimentally. 
As was realized in the experiment, we now fix $V_0$ and monitor the particle behavior as a function of the drive frequency $\W_d/2\pi$. 
We solve \eqnref{eq:Parmeteric_Pendulum} numerically, with librating initial conditions $\alpha(0)= 5\times 10^{-3}~\text{rad}$ and $\Dot{\eula}(0)=0.01~\W_d/2$ and for trap length scale $\ell_0=30~\mu\text{m}$ and radial asymmetry $a_x-a_y = 0.103$. To compute the quadrupole tensor we assumed the particle to be a prolate spheroid with major and minor axes $b$ and $a$ respectively. Accordingly, for $a/b\ll 1 $, we have $\Delta Q\equiv Q_2-Q_3 \simeq q_\text{tot} b^2(1+2a^2/b^2)/4$~\cite{Rusconi2022}. Just as the typical particles used in the experiment, we assumed $b=15~\mu\text{m}$ and $a=4~\mu \text{m}$, $q_\text{tot}=2500~e$, with $e$ the electron charge. From the results of the numerical integration we compute $\eta_\text{rot}$. We then repeat the procedure for different values of $V_0$. 

The results are shown in \figref{fig:Fig3}(a),
 where we marked with a blue dot the value $\W_{l\rightarrow r}$ at which the system switches from the librating to the rotating regime.
The values of $\eta_\text{rot}$ as a function of $\Omega_d/2\pi$ are shown in \figref{fig:Fig3}(b) (red circles). We see that the particle motion departs from a pure libration (where $\eta_\text{rot} = 0$) to a rotation where $\eta_\text{rot} = +1$ at about 4.4 kHz. Then, as indicated by a white region in \figref{fig:Fig3}(a), in the smaller frequency range, the motion is not a pure rotation anymore, and in fact resembles that of a pure libration. This is apparent from the value of $\eta_{\rm rot}$ which becomes closer to zero in the 1.5 to 2 kHz range. There, the motion is extremely sensitive to the parameter values, signalling a potentially chaotic response, as expected for a parametric pendulum at a low driving frequency~\cite{Kapitza1951,VanDerWeele2001,Xu2005,Litak2008}.
Studying this limit in detail however goes beyond the scope of this paper. 

We then proceed to describe the opposite situation, where we start from the rotating regime and increase the drive frequency.
This time the numerical solution is calculated with the initial rotating conditions $\alpha(0)=5\times 10^{-3}~\text{rad}$, $\Dot{\eula}(0) =  1.01~\W_d/2$~\footnote{The case of $\Dot{\eula}(0)=-1.01~\W_d/2$ leads to identical results.}. In \figref{fig:Fig3}.(a), we marked with a red circle the smallest value of $\W_{r\rightarrow l}$ at which we obtain $\eta_\text{rot}=0$. 
We see that $\W_{r\rightarrow l}>\W_{l\rightarrow r}$ systematically. The order parameter $\eta_\text{rot}$ thus exhibits a hysteresis in the region comprised between $\W_{r\rightarrow l}$ and $\W_{l\rightarrow r}$. We note that the dynamical phase diagram obtained theoretically in \figref{fig:Fig3}.(a) does not only recover the three observed phases but captures also the functional dependence of the experimental curves corresponding to $\W_{r\rightarrow l}$ and $\W_{l\rightarrow r}$ [cf. \figref{fig:Fig2}.(c)].
As shown in Fig. 3-c), other rich angular dynamics can take place when the particle angle enters the rotating regime. These have not been analysed experimentally so we leave the theoretical analysis in Appendix \ref{misc_regimes}.

Electrical locking regime provides a natural rotation mechanism for particles levitating in Paul traps. It was realized here in a regime where the particles are highly anisotropic. Optical read-out is then straightforward with such large aspect-ratio particles, enabling unambiguous analysis of their angular dynamics. One of our prospects for using such electrical locking effect was however to set the scene for observing gyroscopic effects taking place when crystalline particles contain internal degrees of freedom. Observing the motion and entering the locking regime is then not a trivial task because crystalline particles that contain isolated spins are often irregularly shaped. 
In the next section, we make use of the spins themselves to probe the motion of rotating crystalline particles. Specifically, we trap diamond particles containing NV centers and use NV magnetometry to measure the particle rotation. 

\section{Angular Motion Readout using NV centers}\label{sec:Rotate_NV}

NV centers in diamonds are widely employed in magnetometry because of the possibility to polarize and read out their electronic spins under ambient conditions. NV centers inside a freely moving diamond can also serve as a probe of the particle motion by monitoring the change in the photoluminescence in a known magnetic field as the particle is moving. This technique has already been employed to characterize the angular stability of levitated diamonds in the librating regimes of Paul traps~\cite{Delord2017APL} or, more recently, for the 6D tracking of a moving biological membrane using a tethered nano-diamond~\cite{Feng2021}. 
Here, we demonstrate tracking of the angular trajectory of a levitating diamond in the rotational-locking regime using NV centers. 

\subsection{Rotational-locking of diamond micro-particles}


\begin{figure}
	\includegraphics[width=\columnwidth]{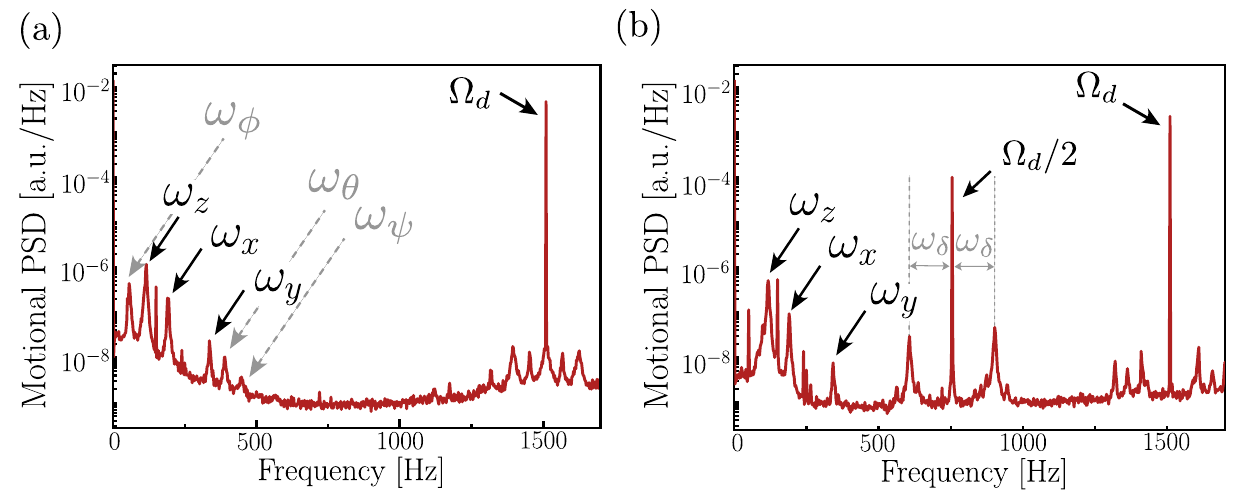}
	\caption{Power spectral density of the motion of a levitating diamond at the pressure $P=0.7~\text{mbar}$ in the librating regime in (a), and in the rotational-locking regime in (b) with $\W_d/2\pi=1510~\rm{Hz}$. The same diamond particle is employed in both measurements. }\label{fig:Fig2_bis}
\end{figure}

In \figref{fig:Fig2_bis}, we present the power spectral densities obtained with an irregularly shaped $15~\mu$m levitated High Pressure High Temperature diamond from Adamas Nanotechnologies (Adamas Nanotechnologies,
Raleigh, NC, USA) at a pressure of $P=0.7~\text{mbar}$. 
Working at this low pressure allows us to fully resolve the center of mass and librational confinement frequencies.
In \figref{fig:Fig2_bis}.(a), we display the PSD of the diamond in the librating regime while, in \figref{fig:Fig2_bis}.(b), we present the PSD of the diamond in the rotational-locking regime where a peak at $(\Omega_d/2)/2\pi$ appears. The two PSDs were obtained with the exact same Paul trap parameters $\Omega_d/2\pi=1510~{\rm Hz}$ and $V_0=400~{\rm V}$. We used the previously explained hysteretic behavior of the angular regime to obtain these two PSDs. We ensured that the diamond was rotating by using the AOM stroboscopic detection scheme depicted in \figref{fig:Fig2}.(b).

Comparing the PSDs in these two regimes allows for the interpretation of some of the observed peaks as center of mass or librational modes. The three annotated resonant modes $\w_x,\w_y,\w_z$ appear in both regimes at the exact same frequencies. We thus interpret them as the three center of mass frequencies of the diamond in the Paul trap. To further confirm this interpretation we observed that, as expected, $\w_x,\w_y,\w_z$ decrease for increasing values of $\W_d$ (see \appref{app:MechanicalModes})\cite{Paul1990}. The fact that the center of mass peaks are independent on the rotational motion also suggests that the dipole moment of the particle -- responsible for the coupling between center of mass and rotation~\cite{Martinetz2021} -- is negligible.

The three annotated peaks $\w_\phi, \w_\theta, \w_\psi$ in \figref{fig:Fig2_bis}.(a) vanish when the peak at $\Omega_d/2$ appears [\figref{fig:Fig2_bis}.(b)]. This is indicating that these three modes correspond to librational modes of the particle, thus deeply within the librating regime. We also observe that $\w_\phi, \w_\theta, \w_\psi$ decrease for increasing values of $\W_d$ which is expected from theory (see \appref{app:MechanicalModes}) \cite{Paul1990}. In the rotational-locking regime, two of the angular modes are proportional to $\W_d/2$ because of the gyroscopic stabilization, which differs substantially from the scaling of the libration modes in the librating regime (\appref{app:Theory}). 
Finally we observe sidebands around the peak at $\W_d/2$. The value of the shift $\w_\delta$ is comparable to, but not exactly equal to $\w_z, \w_x,\w_y$. This peaks could thus correspond to a librating mode in the rotating frame, such as one of the two gyroscopic modes, but it is not captured by the theory. Further investigations are needed to clearly identify its physical origin.

\begin{figure*}[t]
\centering
 \includegraphics[width=\linewidth]{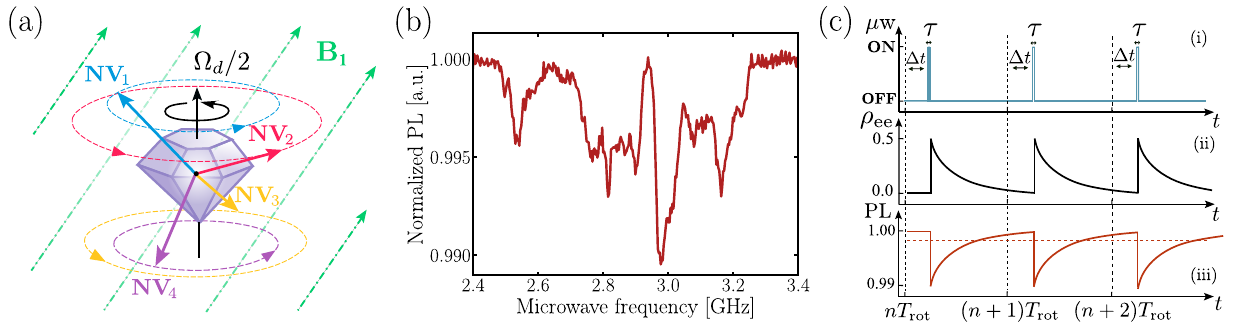}
\caption{(a) Schematics showing a levitating diamond in the rotational-locking regime. NV centers in the four different orientations are depicted. An external magnetic field is applied at a non-zero angle with respect to the diamond rotation axis. (b) Continuous ODMR spectra of a rotating diamond. (c) Experimental sequence used to realize stroboscopic ODMR. (i) shows the sequence applied after the microwave generator where $\Delta t$ is the stroboscopic delay, $\tau$ is the microwave pulse duration. (ii) shows the evolution of the population $\rho_{\text{ee}}$ in the $\ket{m_s=-1}$. (iii) shows the evolution of the photoluminescence (red line) as well as the measured mean value of the photoluminescence (red dashed line).}
\label{fig3}
\end{figure*}

The presence of a sharp peak at $(\Omega_d/2)/2\pi$ in addition to the visualization of the rotation motion using a stroboscoped laser constitute a robust proof of the rotation of the diamond. However, it does not give clear information of the angular dynamics in the rotating frame, and notably whether or not the Euler angles are confined. In the next section, we employ NV magnetometry to demonstrate rotation of the diamond around a single axis, full stability of the three Euler angles in the rotating frame, as well as the remarkable stability of this rotational motion over time. 

\subsection{Angular detection of the diamond rotation using NV magnetometry}

The NV$^-$ center is a point defect in diamond consisting in the association of a nitrogen and a vacancy~\cite{DOHERTY20131}. 
The diamonds we use  contain $3.5~\rm{ppm}$ of NV$^-$ centers which are all equally distributed among the four diamond $[111]$ orientations. The remarkable property of the negatively charged NV$^-$ centers (NV$^-$ center for short) is that its spin can be polarized optically under ambient conditions in the electronic ground state.  
It is an effective spin 1 system in the ground state manifold. The two $\ket{m_s=\pm 1}$ states  are located $D \approx (2\pi) 2.87~\rm{GHz}$ above the $\ket{m_s=0}$ state. The NV$^-$ center can be optically polarized to the $\ket{m_s=0}$ ground state using a green laser. A magnetic field lifts the degeneracy between the two NV$^-$ excited states, making them  separately addressable through resonant microwave excitation. The magnetic state of the NV$^-$ center can be read out by measuring the emitted  photoluminescence (PL), which decreases when an excited state $\ket{m_s=\pm 1}$ is populated. The eight magnetic resonances (two magnetic resonances $\ket{m_s=0} \to \ket{m_s \pm 1}$ for each of the four NV classes) can thus be read out by collecting the PL, while scanning a microwave frequency with a constant green laser illumination. This technique is called Optically Detected Magnetic Resonance (ODMR). 
The value of the transition energies provides direct access to the strength and the direction of the magnetic field in a few milliseconds, making NV$^-$ centers particularly attractive vectorial magnetometers. Here, we make use of the NV$^-$ centers inside the rotating diamond to read out of the diamond angular position.



In \figref{fig3}.(a), we represent  a rotating diamond in an external magnetic field $\abs{\BB_1} \approx 10~{\rm mT}$. The four NV center anisotropy axis rotates in the laboratory frame at a frequency $(\Omega_d/2)/2\pi$, in the kHz range. The projection angle $\theta_1^{(i)}(t)$ between the magnetic field $\BB_1$ and the $i$-th NV center class ($i \in [1,4]$) is a periodic function of time for a rotating diamond and can be written as:
\be\label{eq:theta_i_def}
	\theta_i^{(i)}(t)=\arccos{\left(a_i+b_i\cos{\left(\frac{\Omega_d}{2} t+\phi_i\right)}\right)},
\ee
where $a_i,b_i,\phi_i$ are constant parameters. In order to get information on the angular dynamics of the rotating diamond, we perform continuous ODMR measurement on a rotating diamond. To do so, we continuously illuminate the diamond with the green laser to polarize the $\ket{m_s=0}$ state. A microwave is scanned from $2.4~{\rm GHz}$ to $3.4~{\rm GHz}$ while the NV photoluminescence is detected. The microwave is generated by a microwave generator (Rohde and Schwarz SMB100A) and is brought onto the Paul trap using a bias tee. The microwave frequency is changed every $10~{\rm ms}$, which is ten times larger than the rotation period of the diamond. The diamond thus performs multiple turns at a given microwave frequency. 

The result of this experiment is shown in \figref{fig3}.(b). The observed ODMR spectra features two main regions with decreases PL ranging from $2.45 \to 2.9~{\rm GHz}$ and from $2.9 \to 3.25~{\rm GHz}$. These regions are separated by the avoided crossing between the $\ket{m_s=-1}$ and $\ket{m_s=+1}$ states at $2.9~{\rm GHz}$. Many such spectra have been observed with similar shapes, often showing between four to eight similarly broad dips on both sides of the avoided crossing.
Notably, it does not display the eight 6-8 MHz wide dips which are smoking gun of an angularly stable diamond \cite{Delord2020}.
Further, the ODMR spectra does not correspond to the typical ODMR spectra of a diamond following an angular random walk. All frequencies within the two broad regions on both sides of the avoided crossing would otherwise carry almost equal weight within the frequency range allowed by the magnetic field. This was for instance observed in \cite{horowitz2012electron} for nano-diamonds trapped in liquid. The ODMR we observe may however be consistent with a rotating motion. %

\begin{figure*}
\centering
 \includegraphics[width=\linewidth]{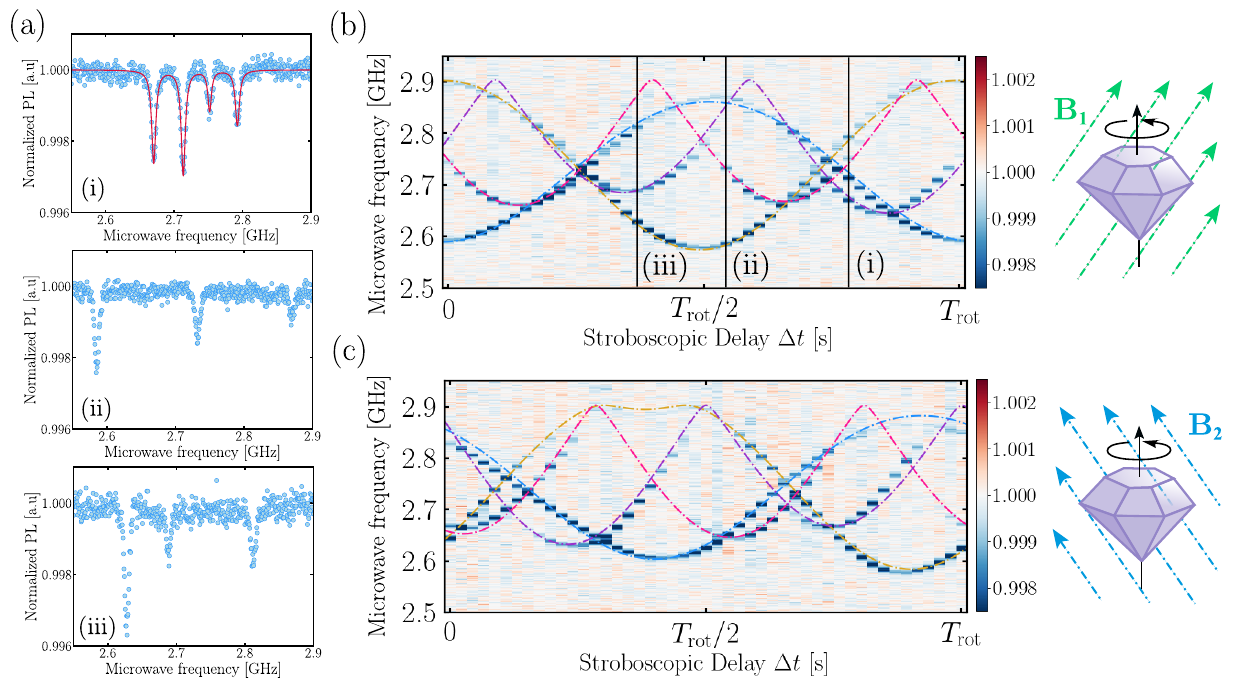}
\caption{(a) (i) Stroboscopic ODMR measurement of the $\ket{m_s=0}$ to $\ket{m_s=-1}$ transitions of the rotating diamond with a delay $\Delta t = 540~\mu{\rm s}$. The red curve corresponds to Gaussian fits of the four transitions with linewidths between $6-7~{\rm MHz}$. (ii)-(iii) Same as (i) but for the delays $\Delta t=375~\mu{\rm s}$ and $\Delta t=255~\mu{\rm s}$ respectively. (b) 2D map of stroboscopic ODMR measurements showing the normalized NV centers photoluminescence as a function of the microwave frequency and the stroboscopic delay $\Delta t$ performed with the magnetic field $\BB_1$. The three vertical black lines correspond to the delay at which the stroboscopic ODMR of (a) were performed. The colored dashed lines are fits corresponding to an ideal rotation at the angular frequency $\Omega_d/2$. (c) Same as (b) with a different magnetic field $\BB_2\approx 10 mT$ perpendicular to $\BB_1$. The colored dashed lines correspond to a fit assuming the same ideal rotation as for (b). The acquisition took approximately four hours of averaging, using $48$ stroboscopic delay values. The contrast of the NV center resonances is less pronounced for microwave frequencies between $2.7-2.9~\rm{GHz}$ than between $2.5-2.7~\rm{GHz}$ due to spurious reflections in the cables. }
\label{fig3_bis}
\end{figure*}

Extracting quantitative information about the diamond rotational dynamics is not trivial from the above continuous ODMR. An alternative would be to 
perform fast ODMR scans in order to track all NV resonances as their frequency change with the diamond rotation. This would however require spectrum acquisitions on timescales on the order of the rotation period ($\approx 100~\mu{\rm s}-1~{\rm ms}$). Unfortunately the sensitivity of the ODMR is not good enough to read-out the magnetic field at such speeds. 

We overcome this difficulty by taking advantage of the diamond rotation periodicity, designing the stroboscopic ODMR sequence depicted in Fig.~\ref{fig3}.(c). 
A fixed frequency microwave pulse of duration $\tau$, synchronised to the Paul trap drive, is periodically switched on at times  $t_n=\Delta t + n T_{\rm{rot}}$ for a given stroboscopic delay $\Delta t$, as shown in \figref{fig3}.(c)-(i). The duration $\tau$ is chosen to be much smaller than the rotation period $\tau/T_{\rm{rot}} =0.1\% $, so that the diamond can be considered as fixed angularly during that time. When the microwave frequency is resonant with an NV$^-$ transition, the population in the $\ket{m_s=0}$ state is transferred to an excited state. There is thus an increase of the population in one of the excited states $\rho_{\text{ee}}$, which then decays with a characteristic timescale of about $10~\mu\text{s}$, given by the optical pumping efficiency of the green laser as depicted in \figref{fig3}.(c)-(ii). The emitted photoluminescence, represented by the continuous red curve in \figref{fig3}.(c)-(iii), decreases when the excited state is populated resulting in a total decrease of the averaged photoluminescence signal (red dashed line). When the microwave frequency is not resonant, the PL remains unaffected. An important point is that the Rabi frequency has to be at least on the order of $1/\tau \approx 1~\text{MHz}$ in order to efficiently address a spin transition during the duration $\tau$. This is ensured experimentally by increasing the signal by 20dBm (using ZHL-5W 422 from Minicircuit).

The microwave frequency is then scanned at a fixed stroboscopic delay $\Delta t=540~\mu{\rm s}$, while measuring the PL from the NV center. 
We typically obtain ODMR spectra such as the one showed in~\figref{fig3_bis}.(a)-(i) within about five minutes, when zoomed on the four $\ket{m_s=0} \to \ket{m_s=-1}$ transitions. Four resonance lines related to the four $\ket{m_s=0} \to \ket{m_s=-1}$ transitions of the NV center classes can be seen. This testifies that the angular position of the diamond is always the same after each round with period $T_{\text{rot}}$. The linewidth obtained from a Gaussian fit equals $1/T_2^* \approx 6-7~\rm{MHz}$, which corresponds to the typical ODMR linewidth observed for NV centers in these diamonds. 

Keeping the magnetic field $\BB_1$ the same, we can now perform stroboscopic ODMR measurements for different values of the stroboscopic delay $\Delta t$ between the microwave signal and the Paul trap drive, in the range of $[0,T_{\rm rot}]$. In~\figref{fig3_bis}.(a)-(ii),(iii), we present stroboscopic ODMR measurements for two other values of the delay $\Delta t=375~\mu{\rm s}$ and $\Delta t=255~\mu{\rm s}$. We see that the energy of the $\ket{m_s=0} \to \ket{m_s=-1}$ transitions are shifted. Scanning the delay $\Delta t$ in $[0,T_{\rm rot}]$, we can then obtain a full set of stroboscopic measurements. The result of this experiment is presented in~\figref{fig3_bis}.(b), where the normalized photoluminescence of the diamond is plotted as a function of the microwave stroboscopic delay. The photoluminescence drops corresponding to the NV resonances are the deep blue regions on the graph. We observe that the magnetic resonance frequencies of the four NV classes clearly evolve according to a continuous angular motion at a period $T_{\rm rot}$. Furthermore, precise knowledge of the four NV resonance energies allows us to determine the four angles $\theta_1^{(i)}(\Delta t)$ between each of the four NV classes $i \in [1,4]$ and the direction of the magnetic field for each value of $\Delta t$. 

The knowledge of these four angles is not sufficient to fully reconstruct the angular trajectory of the diamond in the laboratory frame. This is because the values of the angles $\theta_1^{(i)}(\Delta t)$ would remain unaffected by a rotation of the diamond around the magnetic field direction. Consequently, there is still one angular degree of freedom that cannot be determined with the use of a single magnetic field. This lack of knowledge on the angular trajectory can be resolved by performing the same experiment with a second magnetic field $\BB_2$ that is not aligned with $\BB_1$. Here, we chose $\BB_2$ to be perpendicular to $\BB_1$. The results are presented in~\figref{fig3_bis}.(c). As expected, the NV resonances are modified by the change in magnetic field direction. This allows us to obtain a second set of angles $\theta_2^{(i)}(\Delta t)$ between the NV axis and the magnetic field $\BB_2$, from which we can finally deduce the angular trajectory of the diamond.

The evolution of the stroboscopic ODMR resonance lines in~\figref{fig3_bis}.(b) and (c) are fitted by diagonalizing the NV hamitlonian assuming a perfect rotational motion around a single axis at the angular frequency $\Omega_d/2$. There is excellent agreement between the experimental data and the fits, indicating that the diamond is rotating at the frequency $\Omega_d/2$ and that the three rotational modes of the diamond are fully confined in the rotating frame. The angular confinement is here due to the combined action of the quadrupole potential of the Paul trap and the gyroscopic angular stabilization (see \appref{app:Theory}). Only a few milliradians of angular shift have been observed during eight hours of averaging, demonstrating the extreme stability of this rotation technique over time.
This result provides bright prospects for further studies of gyroscopy with NV centers as well as with other magnetic particles.

\section{Discussion and perspectives}

Rotating particles with internal spins offers a plethora of interesting avenues besides the demonstrated motional read-out. 
One area where electrical rotation can be used is for detecting dynamical or geometric phases ~\cite{Maclaurin2012,Chen2019,Wood2020}. Currently, all experiments operate with tethered diamonds where only kHz rotations are currently attained. Larger rotation frequencies could in principle be observed with electrical locking without the typical technical mechanical noise coming instabilities of rotor axes, providing a possibility to bridge the gap between rotation frequency and electronic spin decoherence rate.
Note that the rotation frequencies we attained are already of the right magnitude for observing gyroscopic effects on nuclear spins~\cite{wood2017magnetic}. Nuclear spins are indeed much more isolated from magnetic noise then their electronic counterpart so that their magnetic resonance linewidths lie in the kHz range. It would also offer the tantalizing prospect of using magic angle spinning for reducing the nuclear spin linewidth further, as is routinely done in nuclear magnetic resonance. 

Rotating particles in the MHz range would offer a broader range of applications.
One direction where this could be beneficial is in the field of spin-mechanics.
A previous study was realised with angularly confined diamond with the same NV density \cite{Delord2020}. There, the magnetic torque from the spins was able to displace the angle of diamonds by about 100$~\mu$rad, as well as to cool down the diamond libration by a factor of four from room temperature.
The ultimate limitation to the cooling efficiency was the low frequency of the mechanical oscillator ($\approx$kHz) compared to the electronic spin transition linewidth ($\approx$10 MHz). Such a large difference between the two systems prevented entering the so-called sideband resolved regime where anti-Stokes heating is mitigated. One solution to bridge this large frequency gap could be to rotate the diamond particle and benefit from the gyroscopic stability of the angular modes in the co-rotating frame that is directly related to the rotational frequency.

In principle, the maximum achievable locking frequency using our method is limited by the damping rate resulting from collisions with the background gas (see Appendix \ref{Dissipation}). According to \eqnref{eq:nu_max}, it should be possible to reach rotation rate in the $\text{MHz}$ range at a pressure of $10^{-1}~\text{mbar}$. The same rates are obtained at a similar pressure with optical rotation using tweezers~\cite{Reimann2018}. Although tweezers are primarily utilized for spinning much smaller particles, rotational frequencies in the MHz range for $10~\mu\text{m}$ particles have been demonstrated~\cite{Monteiro2018}. However, this technique has not been applied to rotate magnetic particles such as diamonds, which rapidly heat up at low pressures due to light absorption. Electric rotation using the Paul trap rotational-locking regime circumvents this issue. Currently, the primary practical limitation in achieving such a regime in our setup is the loss of the particle at frequencies above the kHz range. At higher frequencies, the stability region for confining the center of mass becomes narrower, making it more challenging to retain the particle in the trap. However, compensating for the micromotion caused by the gravitational force, combined with feedback cooling of the three center of mass modes, should be sufficient to stabilize the particle's position, even when the resonant frequencies of the center of mass modes are low. At this point, one could also consider adding a second drive to the Paul trap in the kHz range, which would stabilize the center of mass motion, while the first Paul trap drive in the MHz range would be responsible for the rapid rotation of the particle. 

\section{Conclusions}\label{sec:Conclusion}
In conclusion, we experimentally demonstrated an \emph{all-electric} protocol to rotate microparticles in a standard Paul trap up to few kHz at pressures ranging from atmospheric pressure down to one millibar. 
This method is based on the intrinsic non-linearity for the angular dynamics of an object in a Paul trap.
Specifically by controlling the parameter of the trap such as voltage amplitude and frequency it is possible to switch to different dynamical regime for the particle's rotational motion. As such, the method is extremely versatile and it can be applied to a large variety of particles like ferromagnets. 
We indeed could use diamonds with embedded color centers to realize NV magnetometry and to reconstruct the particle angular trajectory, thereby demonstrating the single-axis character of the rotation mechanism. 

Our results are first step towards precise control of fast rotating particles with an internal magnetic structure, thus opening the door to the experimental investigation of the interplay between orbital angular momentum and spin angular momentum with macroscopic particles. A particularly intriguing new research direction is the search for atomic-like effects on magnet motion, stemming from by the spin degree of freedom \cite{Kimball2016, Rusconi2017,Kustura2022}. Observing such effects is under reach using trapped nano-ferromagnets or particles containing a large number of spins and could lead to several applications in gyroscopy, magnetometry \cite{Vinante2021}, spin-mechanics \cite{Huillery2020, Gieseler2020}, or in fundamental tests of quantum mechanics \cite{Timberlake2021,Rusconi2022}. 
While the observation of some of these phenomena requires rotation rates comparable or larger than spin-dephasing rates ($\sim$MHz), we see no fundamental limitation in reaching higher rotational frequencies with our method.


\begin{acknowledgments}
We thank Haggai Landa and Oriol Romero-Isart for stimulating discussions. B.A.S. acknowledges support by the Deutsche Forschungsgemeinschaft (DFG, German Research Foundation) -- 510794108. M.P.\ and G.H.\ have been supported by Region Île-de-France in the framework of the DIM SIRTEQ.
This project was funded within the QuantERA II Programme that has received funding from the European Union’s Horizon 2020 research and innovation programme under Grant Agreement No 101017733.
\end{acknowledgments}

\appendix

\section{Theoretical Description of an electrically levitated rotor in a Paul trap}\label{app:Theory}

In this appendix, we model the three dimensional dynamics of an electrically levitated rotor in a Paul trap. We show how, from this general description, one can derive the simple model in \eqnref{eq:Parmeteric_Pendulum}, and the characteristic frequencies of the secular oscillations around the librating and rotating solutions.

We model the levitated particle as an asymmetrical rigid body. The degrees of freedom of the system are thus its center of mass position $\RR$ and its orientation in space. This latter is parameterized by the generalised coordinates $\W$.
The dynamics of the system can then be described by the following set of equations
\begin{subequations}
\bea
    M\Ddot{\rr} &=& \FF(\RR,\W,t),\label{eq:EoM_CM}\\
    \Dot{\LL} &=& \NN(\RR,\W,t).\label{eq:EoM_Rotation}
\eea
\end{subequations}
Here, $M$ is the mass of the rotor, $\LL$ is the rigid body angular momentum, and $\FF(t,\RR,\W)$ [$\NN(t,\RR,\W)$] is the time dependent force (torque) exerted by the Paul trap potential.
They can be obtained from the potential energy of the particle in the trap. This is calculated by integrating the surface charge distribution $\varrho(\rr)$ over the trap potential \eqnref{eq:V_p}. Due to the quadrupole symmetry of the potential we obtain
\be
    U(\rr,\W,t) = \frac{V(t)}{\ell_0^2} \Big[ q {\bf R} \cdot {\sf A} {\bf R} + 2 {\bf p}(\Omega) \cdot {\sf A} {\bf R} + \frac{1}{3} \Tr\left [ {\sf Q}(\Omega) {\sf A} \right ] \Big].\nonumber
\ee
where $q$ is the total surface charge and we introduced the tensor
\be
    {\sf A} \equiv a_y{\bf e}_y \otimes {\bf e}_y + a_x{\bf e}_x \otimes {\bf e}_x  +a_z {\bf e}_z \otimes {\bf e}_z.
\ee
We also define the particle dipole moment  $\pp(\W) = {\sf R}(\W) \pp_0$, and quadrupole tensor ${\sf Q}(\W) = {\sf R}(\W) {\sf Q}_0 {\sf R}^T(\W)$, where
\begin{subequations}
\bea
    \pp_0 &\equiv& \int_S\!\! d\rr \,\varrho(\rr) \, \rr,\label{eq:p_0} \\
    {\sf Q}_0 &\equiv& \int_S\!\! d \rr \,\varrho(\rr) \, \Big( 3 \rr \otimes \rr - r^2 \mathbb{1} \Big)\label{eq:Q_0}
\eea
\end{subequations}
are the dipole moment and quadrupole tensor defined with respect to the particle center of mass.
The transformation ${\sf R}(\W)$ relates the laboratory-fixed frame $\Olab$ and the body-fixed frame $\Obod$ according to $\nn_k= {\sf R}(\W)\bold{e}_k$. In the following, we will parameterize the particle orientation with the Euler angles $\W=(\eula,\eulb,\eulc)$  according to the  $zy'z''$ convention. Within this choice the rotation matrix reads
\be\label{eq:R_euler}
\begin{split}
    {\sf R}(\W)\equiv& \begin{pmatrix}
		\cos \alpha & -\sin \alpha & 0\\
		\sin \alpha & \cos \alpha & 0\\
		0 & 0 & 1
	\end{pmatrix}
	\begin{pmatrix}
		\cos \beta & 0 & \sin \beta \\
		0 & 1 & 0\\
		-\sin \beta & 0 &  \cos \beta \\
	\end{pmatrix}\\
	&
	\begin{pmatrix}
		\cos \gamma & -\sin \gamma & 0\\
		\sin \gamma & \cos \gamma & 0\\
		0 & 0 & 1
	\end{pmatrix}.
\end{split}
\ee
In our experiment, we never observed a coupling between the center of mass and rotational dynamics. This suggest that our particle have a negligible dipole moment and we thus assume $\pp_0=0$. 
In this case, \eqnref{eq:EoM_CM} and \eqnref{eq:EoM_Rotation} can be treated independently.
In the following we consider only the rotational dynamics of the particle.

Let us now express the rotational dynamics of the particle in terms of the Euler angles coordinates introduced above. 
In the body-fixed frame $\Obod$, we can express the angular momentum as $\LL = {\sf I} \boldsymbol{\w}$, where ${\sf I} \equiv \sum_k I_k \nn_k\otimes \nn_k$ and $I_k$ ($k=1,2,3$) are the constant principal moment of inertia.
Substituting this expression into \eqnref{eq:EoM_Rotation} we obtain the well known Euler equations
\be\label{eq:Euler_Eq}
\begin{split}
    I_1 \Dot{\w}_1 - (I_2-I_3)\w_2\w_3 =& N_1(\W,t),\\
    I_2 \Dot{\w}_2 - (I_3-I_1)\w_1\w_3 =& N_2(\W,t),\\
    I_3 \Dot{\w}_3 - (I_1-I_2)\w_2\w_1 =& N_3(\W,t),
\end{split}
\ee
where the angular frequencies are related to the Euler angles by
\be\label{eq:w_vect_body}
	\vect{\w_1}{\w_2}{\w_3}\! =\! \begin{pmatrix}
	-\cos\eulc \sin\eulb & \sin\eulc & 0\\
	\sin\eulb \sin\eulc & \cos\eulc & 0\\
	\cos\eulb & 0 & 1
	\end{pmatrix}
	\vect{\deula}{\deulb}{\deulc},
\ee 
and the torque is given by $N_i(\W,t)\equiv \nn_i\cdot \NN(
W,t)$, where $\NN(\W,t)$ is given by\eqnref{eq:Torque}.
Using these results one can show that \eqnref{eq:Euler_Eq} can be obtained from the following Lagrangian
\be\label{eq:Lagrangian}
    \mathcal{L}(\W,\Dot{\W}) \equiv \inv{2}\sum_{k=1}^3 I_k\w_k^2 - \frac{V(t)}{3\ell_0^2}\Tr[{\sf Q}(\W){\sf A}].
\ee

\subsection{Secular Dynamics and equilibrium oscillations}

Let us now compute the dynamics of small oscillations around the librating and rotating solutions to \eqnref{eq:Euler_Eq}.
In the general case, the dynamics of such oscillations is complicated by the parametric driving. 
We are interested in isolating the secular component of these oscillations. In the regime where this is possible, the secular oscillations are harmonic and have well defined secular frequencies. In the following we consider separately the case of librating and rotating regime.

\begin{figure}
    \centering
    \includegraphics[width=\columnwidth]{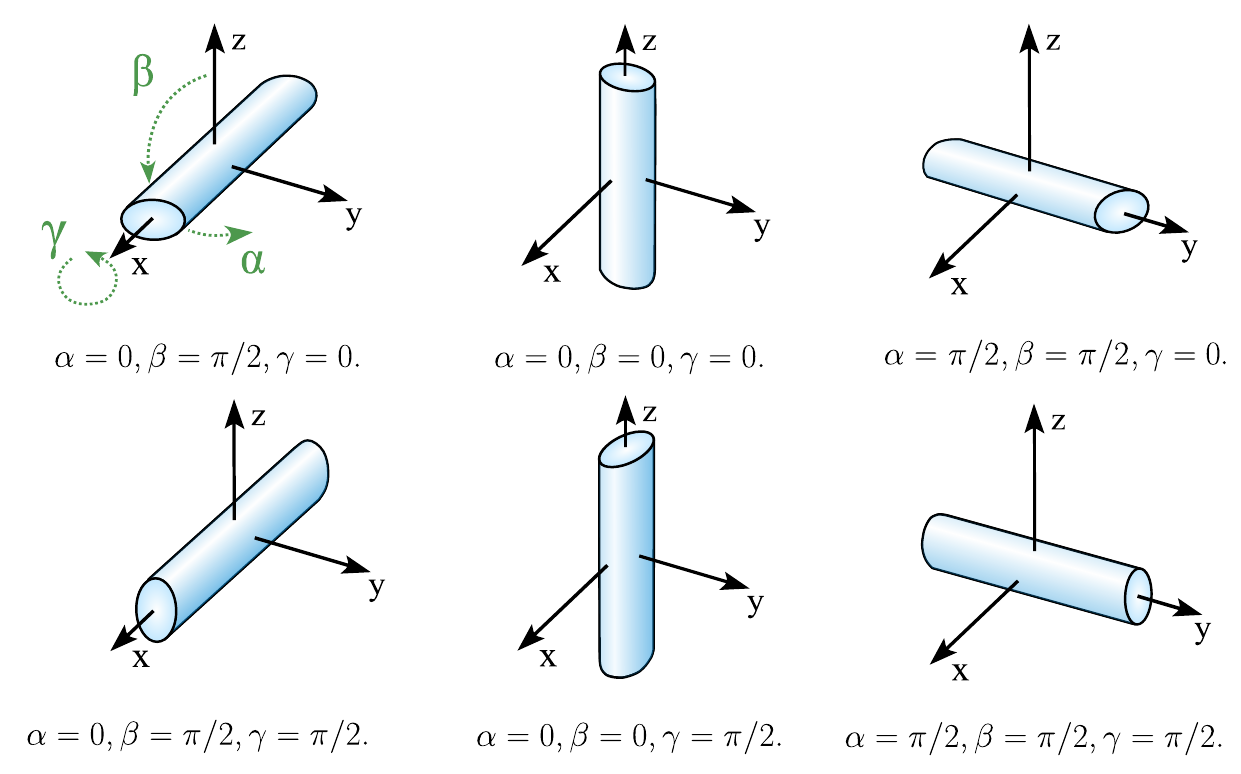}
    \caption{Different equilibrium orientation for an asymmetric particle in the Paul trap potential. In all the cases we have $\deula=\deulb=\deulc=0$.}
    \label{fig:Fig_Equilibria}
\end{figure}

The librating regime of a charged rotor in a Paul trap has been described in~\cite{Martinetz2021}. In particular, the general form of the effective potential for the secular dynamics of the rotor is given by Eq.(53) in~\cite{Martinetz2021}. Analysing this potential for our situation, we find six distinct stable equilibrium positions illustrated in \figref{fig:Fig_Equilibria}.
Let us consider the equilibrium solution at $\eula=0,\eulb=\pi/2$ and $\eulc=0$, that corresponds to the librating solution in the plane where the locking occurs. This is the librating regime to which the particles returns after the hysteresis cycle shown in~\figref{fig:Fig3}.b. Linearizing about this regime, we obtain that the harmonic fluctuation of the three angles are decoupled, and have the following characteristic frequencies
\begin{subequations}
\bea
    \w_\eula\! &\equiv&\! \frac{2}{I_1}\pare{\frac{V_0}{3\ell_0^2 \W_d}}|\pare{a_x-a_y}\pare{Q_2-Q_3}|,\label{eq:w_alpha_lib}\\
    \w_\eulb\! &\equiv&\! \frac{2}{I_2}\pare{\frac{V_0}{3\ell_0^2\W_d}}|\pare{a_x-a_z}\pare{Q_1-Q_3}|,\label{eq:w_beta_lib}\\
    \w_\eulc\! &\equiv&\! \frac{2}{I_3}\pare{\frac{V_0}{3\ell_0^2 \W_d}}|\pare{a_y-a_z}\pare{Q_1-Q_2}|.\label{eq:w_gamma_lib}
\eea
\end{subequations}
We remark that the separation between macromotion and micromotion that underpin the derivation of \eqnref{eq:w_alpha_lib}-\eqnref{eq:w_gamma_lib} is valid only when the following condition is satisfied (for $i,j,k=1,2,3$)
\be\label{eq:Secular_Condition}
    \frac{V_0 (Q_i-Q_j)}{I_k \W_d^2 \ell_0^2}\ll1.
\ee
When this is not the case, such as in the regime we operate our experiment, the separation between macromotion and micromotion is not rigorously possible. We expect, however, Eq.~(\ref{eq:w_alpha_lib}-\ref{eq:w_gamma_lib}) to capture the dependence of the libration peaks observed in the PSD of our experiment.

Let us now consider fluctuation around the rotational-locking solution. To derive the frequency of these fluctuations it is convenient to move to a frame co-rotating with the particle. This is done making the substitution $\eula \rightarrow \eula + \W_d t/2$ in \eqnref{eq:Lagrangian}. The Lagrangian in the co-rotating frame reads
\be\label{eq:Lagrangian_rot}
\begin{split}
    \mathcal{L}'(\W,\Dot{\W}) =& \inv{2}\sum_{k=1}^3I_k\!\pare{\w_k+\frac{\W_d}{2}\bold{e}_z\cdot \nn_k}^2\!\! -\! U_0\! -U_1(t).
\end{split}
\ee
We separated the quadrupole potential into two terms. The first term, $U_0$, is time-independent and reads
The 
\be
\begin{split}
    U_0 =& \frac{V_0}{12\ell_0^2}\pare{a_x-a_y}\Big\{\Big[Q_2\pare{\cos^2\eulb \sin^2\eulc-\cos^2\eulc}\\
    &+Q_1\pare{\cos^2\eulb\cos^2\eulc-\sin^2\eulc}\Big]\cos2\eula\\
    &+\pare{Q_2-Q_1}\cos\eulb\sin2\eulc\sin2\eula\\
    &+Q_3\sin^2\eulb\cos2\eula\Big\}.
\end{split}
\ee
The second term, $U_1(t)$ is a time dependent potential and reads
\be
\begin{split}
    U_1(t) \equiv& u_1(\W)\cos(\W_d t)+u_2(\W)\cos(2\W_d t)\\ &+u_3(\W)\sin(2\W_d t),
\end{split}
\ee
where we defined the following functions
\begin{widetext}
\bea
    u_1(\W) &\equiv& \frac{V_0}{3\ell_0^2} 
    \Big\{a_z\spare{Q_3\cos^2\eulb +\pare{Q_1 \cos^2\eulc+Q_2\sin^2\eulc}\sin^2\eulb}+\pare{\frac{a_x+a_y}{2}}\Big[\pare{Q_1\sin^2\eulc+Q_2\cos^2\eulc}\nonumber\\
    & &+\{Q_3\sin^2\eulb +\pare{Q_1 \cos^2\eulc+Q_2\sin^2\eulc}\cos^2\eulb\Big]\Big\},\\
    u_2(\W) &\equiv& \frac{V_0}{6\ell_0^2}\pare{\frac{a_x-a_y}{2}}\Big\{\cos(2\eula)\Big[Q_3\sin^2\eulb +\pare{Q_2\sin^2\eulc+Q_1\cos^2\eulc}\cos^2\eulb-\pare{Q_1\sin^2\eulc+Q_2\cos^2\eulc}\Big]\nonumber\\ 
    & & +\pare{Q_2-Q_1}\cos\eulb \sin(2\eulc) \sin(2\eula) \Big\},\\
    u_3(\W) &\equiv & -\frac{V_0}{6\ell_0^2}\pare{\frac{a_x-a_y}{2}}\Big\{\sin (2\eula) \Big[Q_3\sin^2\eulb +\pare{Q_2\sin^2\eulc+Q_1\cos^2\eulc}\cos^2\eulb-\pare{Q_1\sin^2\eulc+Q_2\cos^2\eulc}\Big]\nonumber\\ 
    & & +\pare{Q_2-Q_1}\cos\eulb \sin(2\eulc) \cos (2\eula) \Big\}.
\eea
\end{widetext}
From \eqnref{eq:Lagrangian_rot} we then obtain the equations of motion of the system. 
The secular potential will have both the static contribution of $U_0$ and the additional correction coming from the secular approximation of $U_1(t)$ obtained with the method of \cite{Martinetz2021}.
This latter are much smaller than $U_0$, and we shall thus neglect them.
The dominant contribution to the effective potential arises from the kinetic energy and it is proportional to $(\W_d/2)^2$. It acts only on the angles $\eulb$ and $\eulc$ and represents the gyroscopic confinement produced by the particle rotation. 
Linearizing the equation of motion obtained from \eqnref{eq:Lagrangian_rot} around $\eula=\eulb=\pi/2$ and $\eulc=0$ we have that $\eula$ decouples from the other degrees of freedom and performs harmonic oscillations at the frequency 
\be
    \tilde{\w}_\eula \equiv \sqrt{\frac{V_0 (Q_2-Q_3)}{3I_1 \ell_0^2}\pare{a_y-a_x}}.\label{eq:W_alpha}
\ee
The remaining degrees of freedom evolve instead according to
\bea
    \Ddot{\eulb} &=&-\tilde{\w}^2_\eulb \eulb -\frac{\W_d}{2}\pare{\frac{I_1}{I_2}-1}\Dot{\eulc},\\
    \Ddot{\eulc} &=& -\w_\eulc^2\eulc + \frac{\W_d}{2}\Dot{\eulb},
\eea
where the characteristic frequencies read
\bea
    \tilde{\w}_\eulb\! &\equiv&\!\! \sqrt{\frac{I_1}{I_2}\pare{\frac{\W_d}{2}}^2\!-\!\frac{V_0 (Q_1-Q_3)}{6I_2 \ell_0^2}\pare{a_x-a_y}},\label{eq:W_beta}\\
    \tilde{\w}_\eulc\! &\equiv&\! \bigg[\frac{I_1}{I_3}\pare{\frac{I_1}{I_2}-1}\pare{\frac{\W_d}{2}}^2\nonumber\\
    & &-\frac{V_0 (Q_2-Q_1)}{6 I_3\ell_0^2}(a_x-a_y)\bigg]^{1/2}.\label{eq:W_gamma}
\eea
Let us note that the stable equilibrium solution for $\eula$ is $0$ or $\pi/2$ depending whether $a_x>a_y$ or conversely. Linearising around $\eula=0$ instead than $\eula=\pi/2$, as done here, flips the sign in front of $(a_x-a_y)$.

\subsection{Beyond libration and full rotation}\label{misc_regimes}

Fig.~3-c) shows 3 different regimes that we now analyse in more details. The transition from the librating to the rotating regime at $\W_{l\rightarrow r}$ is explained by a loss of stability of the confined solutions $\eula= k\pi/2$, $\Dot{\eula}=0~\text{rad.s}^{-1}$. We apply the same Floquet methods generally used for solving the Mathieu equation to investigate the stability of \eqnref{eq:Parmeteric_Pendulum}~\cite{Kovacic2018}.
The resulting instability region is marked by the hatched area in \figref{fig:Fig3}.(a).
We see that the border of the unstable regime does not perfectly coincide with the transition to the rotational-locking.
This discrepancy arises from the fact that, close to the stability region, $\eula$ can jump between angles $k\pi/2$. This is shown in~\figref{fig:Fig3}(c)-ii), where a selection of four horizontal dashed lines highlights the stable angles. Additionally, we note that within the stable rotating regime but close to the border of instability, $\eula$ performs large amplitude oscillations around the initial equilibrium value at a frequency locked to the voltage frequency at $\W_d/2$ [see \figref{fig:Fig3}(c)-iii)]. This locked oscillations are detected as a peak at $\W_d/2$ in the PSD and can be mistaken for rotational-locking as discussed in \secref{sec:Experimental_Results}.

\subsection{Effects of Dissipation}\label{Dissipation}

Until now we have considered only the dynamics of the system in the absence of dissipation. 
For the pressure at which the experiment is operated, scattering with background gas particles represents the largest source of dissipation. 
The effects of the viscous drag coming from the background gas can be included as shown in \eqnref{eq:Euler_EoM}. The form of the tensor $\Gamma$ depends on both the shape of the rotor and on the properties of the background gas. In particular, it depends on the ratio between the particle size and the mean free path of the gas. In the Knudsen regime, \ie when the particle size is smaller than the mean free path of the gas, $\Gamma$ can be obtained as shown in~\cite{Martinetz2018}. When this is not the case, as for our experiment, the form of the tensor is not easy to obtain. For the theoretical calculations discussed in \ref{sec:Theory} we assumed a damping rate of $\gamma_0/2\pi=1~\text{kHz}$ which is of the same order of magnitude as observed in the experiment.   \eqnref{eq:Parmeteric_Pendulum} can be obtained from the general model presented here by evaluating \eqnref{eq:Euler_Eq} on $\eulb=\pi/2$ and $\eulc=0$ and adding a phenomenological damping rate $\gamma_0$.

Let us now focus on the transition from rotation to confinement at $\W_{r\rightarrow l}$. The stability of the rotating regime is best investigated in the co-rotating frame at the locking frequency. Hence, we transform \eqnref{eq:Parmeteric_Pendulum} according to $\eula \rightarrow \eula - \W_d t /2$.
We now consider the case $\W_d \gg \w_0$, and after averaging over the period of the micromotion, we obtain the following equation for the secular dynamics in the co-rotating frame 
\be\label{eq:EoM_CoRotating}
	\Ddot{\eula} + \gamma_0 \Dot{\eula} + \frac{\w_0^2}{2}\sin(2\eula) = - \frac{\gamma_0 \W_d}{2}.
\ee
\eqnref{eq:EoM_CoRotating} describes the angular secular dynamics of the particle in the rotating frame.
For the rotation to remain stable, the rotational speed cannot exceed a certain value $\W_\text{max}$  at which the torque induced by gas collisions becomes stronger than the restoring torque of the Paul trap. This establishes an upper-bound condition on the Paul trap drive frequency that is directly lied to the rotational speed as a function of the damping coefficient
\be\label{eq:nu_max}
	\W_d < \W_\text{max} \equiv \frac{\w_0^2}{\gamma_0}.
\ee
As shown by the red dashed line in~\figref{fig:Fig3}.(a), $\W_\text{max}$ roughly approximates $\W_{r\rightarrow l}$ and captures its linear dependency on $V_0$. We note that \eqnref{eq:nu_max} can be obtained rigorously in the adiabatic regime of the Paul trap (\ie when $\w_0/\W_d\ll 1$).
In our case instead $\w_0/\W_d\lesssim 1$, which we believe explains the discrepancy between \eqnref{eq:nu_max} and $\W_{r\rightarrow l}$ in \figref{fig:Fig3}.(a).

\section{Mechanical modes in the librating and rotational-locking regime}\label{app:MechanicalModes}

\begin{figure}[h]
	\includegraphics[width=\columnwidth]{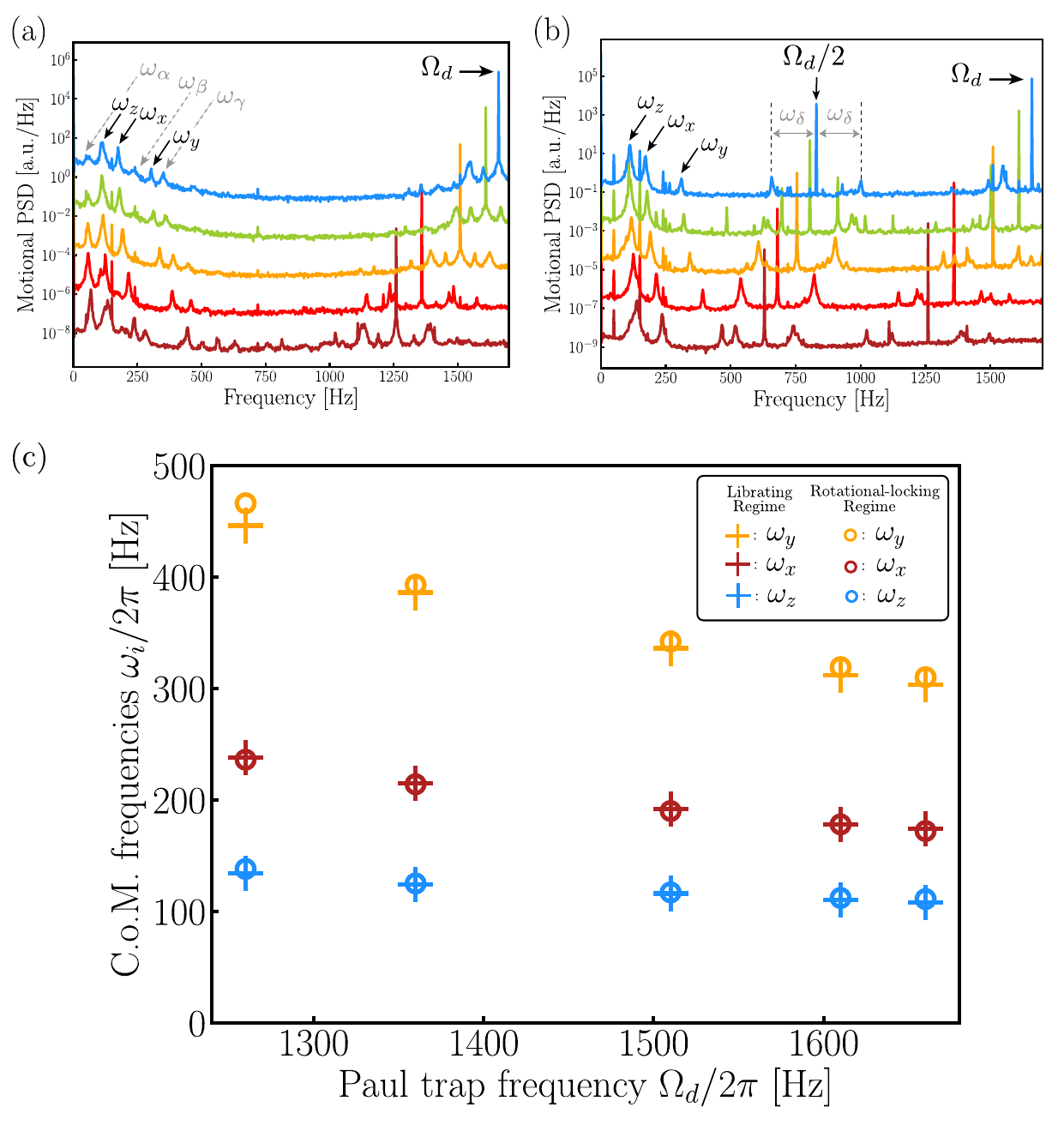}
	\caption{Experimental results: PSDs of a levitating diamond being in the librating regime (a) and in the rotational-locking regime (b) for different drive frequencies: $\Omega_d/2\pi=1260~\text{Hz}$ (purple), $1310~\text{Hz}$ (red), $1510~\text{Hz}$ (yellow), $1610~\text{Hz}$ (green), $1660~\text{Hz}$ (blue). For clarity, the background noise has been shifted by $20$ dBm between each plots. (c) Center of mass frequencies of a levitating diamond for the aforementioned Paul trap drive values.}\label{fig:Fig_PSDs}
\end{figure}

In this appendix, we present extended datas of the two PSDs shown in \figref{fig:Fig2_bis}, where a levitating diamond can be either found in the librating regime or in the rotational-locking regime within the hysteretic angular stability domain. With the same method than the one used to obtain the results in \figref{fig:Fig2_bis}, we obtained different PSDs of the levitating for different values of the Paul trap drive frequency $\Omega_d/2\pi$ in the two different angular regimes. We present these results in \figref{fig:Fig_PSDs}.(a) for the librating regime and in \figref{fig:Fig_PSDs}.(b) for the rotational-locking regime for five different values of the Paul trap frequency drive: $\Omega_d/2\pi=1260~\text{Hz}, 1310~\text{Hz}, 1510~\text{Hz}, 1610~\text{Hz}, 1660~\text{Hz}$. 
The three center of mass (C.o.M) modes have been identified by comparing the PSDs in the two different regimes. In \figref{fig:Fig_PSDs} (c), we have plotted the three C.o.M modes resonant frequencies as a function of the Paul trap frequency drive being in the two different angular regimes. The decrease of the C.o.M. frequencies values with the drive frequency is consistent with Floquet theory. Moreover, the frequencies of the C.o.M. modes do not crucially depend on the angular regime that indicates that there is no coupling between the angular and C.o.M. modes. It justifies that the electric dipole, which could be responsible to a coupling between these modes, can be safely neglected in the calculation.

\end{document}